\documentclass[english,aps,preprint,nofootinbib,showpacs]{revtex4}
\usepackage[T1]{fontenc}
\usepackage[latin9]{inputenc}
\usepackage{amsmath}
\usepackage{amssymb}
\usepackage{esint}

\makeatletter

\providecommand{\LyX}{L\kern-.1667em\lower.25em\hbox{Y}\kern-.125emX\@}

\@ifundefined{textcolor}{}
{%
 \definecolor{BLACK}{gray}{0}
 \definecolor{WHITE}{gray}{1}
 \definecolor{RED}{rgb}{1,0,0}
 \definecolor{GREEN}{rgb}{0,1,0}
 \definecolor{BLUE}{rgb}{0,0,1}
 \definecolor{CYAN}{cmyk}{1,0,0,0}
 \definecolor{MAGENTA}{cmyk}{0,1,0,0}
 \definecolor{YELLOW}{cmyk}{0,0,1,0}
 }

\makeatother

\usepackage{babel}

\begin{document}

\preprint{This line only printed with preprint option}

\title{Some Peculiarities of Newton-Hooke Space-Times}

\author{Yu Tian}

\email{ytian@gucas.ac.cn}


\affiliation{College of Physical Sciences, Graduate University of
the Chinese Academy of Sciences, Beijing 100049, China}

\begin{abstract}
Newton-Hooke space-times are the non-relativistic limit of (anti-)de
Sitter space-times. We investigate some peculiar facts about the
Newton-Hooke space-times, among which the {}``extraordinary
Newton-Hooke quantum mechanics'' and the {}``anomalous Newton-Hooke
space-times'' are discussed in detail. Analysis on the
Lagrangian/action formalism is performed in the discussion of the
Newton-Hooke quantum mechanics, where the path integral point of
view plays an important role, and the physically measurable density
of probability is clarified.
\end{abstract}

\maketitle

\section{introduction}

The Newton-Hooke (NH) space-times \cite{BLL} are very interesting
objects in theoretical physics. On one hand, they are
non-relativistic space-times, which have an absolute time and are
relatively simple. On the other hand, they have non-vanishing
curvature and are akin to the constant curvature space-times, de
Sitter (dS) space-time and anti-de Sitter (AdS) space-time, in many
aspects. Due to their non-relativistic nature, the standard
non-relativistic quantum mechanics can be established on them, which
has been detailedly discussed in \cite{NH} and will be called the
ordinary NH quantum mechanics.\footnote{Although \cite{NH} basically
discusses the case of space dimension $d=3$, it is trivial to
generalize most of the discussions to the case of arbitrary $d$.}
However, due to their non-flat (and even non-static) nature, there
are interesting subtle points in the NH quantum mechanics, which is
intimately related to the Schrödinger group \cite{Sch_group}, as the
{}``conformal'' extension of the NH groups (see e.g. \cite{Galajinsky}), and the Hermiticity of
the NH Hamiltonian. In particular, the NH quantum mechanics is only
established in \cite{NH} by invariance considerations directly on
the Schrödinger equation, while it is not clear whether we can
obtain this equation by some (canonical or path-integral)
quantization procedure. We give thorough clarification of these
points in this paper, starting from analysis on the Lagrangian (or
action) formalism for the NH dynamics and making use of a systematic
path integral point of view. Moreover, it is pointed out that the
probability density defined in \cite{NH} is not NH-invariant and so
cannot be the physically measurable one, while the observable
density of probability in NH space-times is proposed. As a
byproduct, an elegant geometric interpretation of the harmonic
oscillator/free particle correspondence \cite{Niederer} will be
described.

The standard NH space-times as affine connection spaces have also
been discussed in \cite{NH}, with the affine connection uniquely
determined by the NH invariance of both the connection itself and
the affine parameter. In fact, it is taken for granted that the
affine parameter is identified, up to constant linear
transformations, to the NH-invariant proper time $\tau$.
Interestingly, it turns out that consistent mechanics can be
established on affine connection spaces with NH-invariant
connections and a non-invariant affine parameter, which will be
called the anomalous NH space-times. The so-called linear
coordinates on the (anomalous) NH space-times are naturally
introduced there, under which the NH {}``isometry'' transformations
become linear. It is also interesting to see that these coordinates
are related to the Schrödinger group. Furthermore, as an example of
non-trivial dynamics, the Newton-Cartan-like gravity \cite{NH} in
the anomalous NH space-times will be shown in this paper.

This paper is organized as follows. We first briefly review the NH
space-times and NH mechanics in Sec.\ref{sec:NH_Space-Times}, concentrating
on the symmetry and quantum aspects. Then the subtleties in the NH
quantum mechanics are analyzed in Sec.\ref{sec:NH_QM}, where the
so-called extraordinary NH quantum mechanics are proposed. The anomalous
NH space-times are discussed in Sec.\ref{sec:Anomalous_NH} finally.

\section{Newton-Hooke Space-Times and Newton-Hooke Mechanics\label{sec:NH_Space-Times}}

First, we review some relevant facts about the NH space-times and
the mechanics on them. The NH space-times can be described from either
the algebraic or the geometric point of view.

Algebraically, the $(1+d)$-dimensional NH space-times can be constructed
as the homogeneous spaces $N_{\pm}(1,d)/\hat{N}_{\pm}(1,d)$, where
$N_{\pm}(1,d)$ are the NH groups with positive/negative {}``cosmological
constant'' and $\hat{N}_{\pm}(1,d)$ the corresponding homogeneous
NH groups, with the upper/lower sign sometimes called the NH/anti-NH
(ANH) case, respectively. The NH groups are certain non-relativistic
limit (contraction \cite{IW}) of the dS/AdS groups. The Lie algebras
of the NH groups, called $\mathfrak{n}_{\pm}(1,d)$, are described
by the following Lie brackets between anti-Hermitian generators:\begin{eqnarray}
[J_{ij},J_{kl}] & = & \delta_{jk}J_{il}+\delta_{il}J_{jk}-\delta_{ik}J_{jl}-\delta_{jl}J_{ik},\nonumber \\{}
[J_{ij},P_{k}] & = & \delta_{jk}P_{i}-\delta_{ik}P_{j},\qquad[J_{ij},K_{k}]=\delta_{jk}K_{i}-\delta_{ik}K_{j},\nonumber \\{}
[H,P_{j}] & = & \pm\nu^{2}K_{j},\qquad[H,K_{j}]=P_{j},\qquad[P_{i},K_{j}]=0,\label{NH_algebra}\\{}
[J_{j},H] & = & 0,\qquad i,j,k=1,\cdots,d,\nonumber \end{eqnarray}
where $\nu$, sometimes called the NH constant, is the characteristic
of the NH groups (and the corresponding NH space-times). Here, as
usual, $J_{jk}$ are the generators of rotation, $P_{j}$ that of
space translation, $K_{j}$ that of (NH) boosts, and $H$ that of
time translation. When $\nu\to0$, the NH groups just become the Galilei
group. The homogeneous-space structure $N_{\pm}(1,d)/\hat{N}_{\pm}(1,d)$
of the NH space-times can be used to systematically obtain variant
quantities on them, including the so-called {}``canonical'' connection
\cite{KN,ABCP}, from the viewpoint of which the NH space-times can
be regarded as affine connection spaces.

Geometrically, the NH space-times can be directly obtained by the
non-relativistic limit of the dS/AdS space-times, which gives more
straightforwardly the coordinates on the NH space-times and the explicit
NH group actions on them (the so-called NH transformations) as a whole,
thanks to the simple pseudo-sphere nature of the dS/AdS space-times.
Here we omit the limiting procedure and only list some relevant results.%
\footnote{For detailed discussions of the limiting procedure, see \cite{NH_limit,NH}.%
} The NH space-times are of topology $\mathbb{R}^{d+1}$. In order to
explicitly describe their geometry, some suitable coordinate systems
are needed. Two coordinate systems are frequently used in the
literature. One is the so-called Beltrami coordinates $(t,x^{i})$
(with $-\nu^{-1}<t<\nu^{-1}$ in the NH case), which is the
non-relativistic limit of the Beltrami coordinates on the dS/AdS
space-times \cite{BdS}. The other is the so-called static
coordinates $(\tau,q^{i})$, which is the non-relativistic limit of
the well-known static coordinates on the dS/AdS space-times. The
relation between these two coordinate systems is\begin{eqnarray}
\tau & = & \begin{cases}
\nu^{-1}\tanh^{-1}\nu t & \mbox{(for NH)}\\
\nu^{-1}\tan^{-1}\nu t & \mbox{(for ANH)}\end{cases},\label{static}\\
q^{i} & = & \frac{x^{i}}{\sigma(t)^{1/2}},\qquad\sigma(t)\equiv1\mp\nu^{2}t^{2},\nonumber \end{eqnarray}
where $\tau$ is just the NH-invariant proper time (periodic in the
ANH case). The NH transformations under the Beltrami coordinates take
a simple, fractional linear form:\begin{eqnarray}
t^{\prime} & = & \frac{t-a^{t}}{\sigma(a^{t},t)},\qquad\sigma(a^{t},t)\equiv1\mp\nu^{2}a^{t}t,\label{NH_t}\\
x^{\prime i} & = & \frac{\sigma(a^{t})^{1/2}}{\sigma(a^{t},t)}O_{\ j}^{i}(x^{j}-a^{j}-u^{j}t),\label{NH_x}\end{eqnarray}
where $(O_{\ j}^{i})\in SO(d)$ and $a^{t},a^{j},u^{j}\in\mathbb{R}$
are parameters of space rotation, time translation, space translation
and boost, respectively, so these coordinates are more suitable in
this context. It should be emphasized that the transformation for
$t$ is independent of $x^{i}$, so the time simultaneity is absolute,
similar to the Newtonian case. Moreover, the (degenerate) metric\begin{equation}
d\tau^{2}=\sigma^{-2}(t)dt^{2}\label{metric}\end{equation}
is invariant under NH transformations. The generators in (\ref{NH_algebra})
are realized under the Beltrami coordinates as%
\footnote{Summation of repeated indices is assumed throughout this paper, unless
otherwise indicated.%
}\[
H=\sigma(t)\partial_{t}\mp\nu^{2}tx^{i}\partial_{i},\qquad P_{i}=-\partial_{i},\qquad K_{i}=-t\partial_{i},\qquad J_{ij}=x^{i}\partial_{j}-x^{j}\partial_{i}.\]

Taking the Beltrami coordinate systems as inertial (reference) frames,
the NH transformations then act as transformations among inertial
frames, corresponding to the Galilei transformations in Newtonian
mechanics. Differentiation of the NH transformation (\ref{NH_t},\ref{NH_x})
gives rise to the velocity composition law \cite{NH} and the following
transformation of (3-)acceleration: \begin{equation}
\frac{dv^{\prime i}}{dt^{\prime}}=\frac{\sigma^{3}(a^{t},t)}{\sigma^{3/2}(a^{t})}O_{\ j}^{i}\frac{dv^{j}}{dt},\end{equation}
with $v^{i}\equiv dx^{i}/dt$ the velocity. The homogeneous form of
the above transformation means that uniform-velocity motions transform
among themselves under change of inertial frames, implying the existence
of the NH first law of the same form as the Newton's first law. In
fact, it has been shown early that a free point particle moves along
a straight line with uniform velocity in the NH space-time \cite{NH_limit}.
Moreover, the NH second law of the same form as the Newton's second
law also exists provided the force $F^{i}$ has the same NH transformations
as the acceleration. In fact, the NH third law is also consistent
with the NH transformation, and is really respected for the gravitational
interaction on the NH space-times \cite{NH}.

The NH algebra has a unique central extension $\mathfrak{n}_{\pm}^{\mathrm{C}}(1,d)$:\begin{equation}
[P_{i},K_{j}]=-i\delta_{ij}m,\qquad\mbox{and the other Lie brackets same as \ensuremath{\mathfrak{n}_{\pm}(1,d)}},\label{ext_NH}\end{equation}
in generic space dimension $d$, with $m$ a central element.%
\footnote{The case of $d=2$ is very special, where there exists the so-called
exotic central extension. See \cite{exotic,Gao}.%
} The Schrödinger equation on the NH space-times can be deduced by
the algebraic construction from the extended NH algebra
(\ref{ext_NH}) or by the geometric contraction from the Klein-Gordon
equation on the dS/AdS space-times \cite{NH}. Both methods lead to
an NH-invariant equation
\begin{equation}
\mathrm{i}\hbar\partial_{t}\psi(t,x)=[-\frac{\hbar^{2}\nabla^{2}}{2m}\pm\frac{\mathrm{i}\hbar\nu^{2}tx^{i}\partial_{i}}{\sigma(t)}\mp\frac{m\nu^{2}x^{2}}{2\sigma^{2}(t)}]\psi(t,x),\label{Sch_eq}\end{equation}
where we have ignored possible NH-invariant interaction terms. The
conservation of probability for this equation and its relation to
the NH fluid mechanics has been shown \cite{NH}, giving some justification
of the probability interpretation to be physical.

\section{Extraordinary Newton-Hooke Quantum Mechanics\label{sec:NH_QM}}

We base our discussions on the Lagrangian formalism of the NH
mechanics. It is not difficult to show that the non-relativistic
(NH) limit \cite{NH_limit} of the action functional\[ S=-\int
mc^{2}d\tau\] for a free point particle in the dS/AdS space-times
is\begin{equation}
S[x(t)]=\int\frac{1}{2}\frac{m}{\sigma(t)}[\dot{x}^{2}\mp\nu^{2}(x-t\dot{x})^{2}\pm2\frac{\nu^{2}x^{2}}{\sigma(t)}]dt\end{equation}
under the Beltrami coordinates, where an over dot means
$\frac{d}{dt}$. Note that in the process of taking this limit, we
have not discarded any boundary terms. The above action functional
implies a time-dependent Lagrangian\begin{equation}
L=\frac{1}{2}\frac{m}{\sigma(t)}[\dot{x}^{2}\mp\nu^{2}(x-t\dot{x})^{2}\pm2\frac{\nu^{2}x^{2}}{\sigma(t)}]\label{Lagrangian}\end{equation}
for the NH mechanics, which leads to the following canonical momenta
and Hamiltonian:\begin{eqnarray}
p_{i} & = & \frac{m}{\sigma(t)}[\dot{x}^{i}\pm\nu^{2}t(x^{i}-t\dot{x}^{i})]=m[\dot{x}^{i}\pm\frac{\nu^{2}tx^{i}}{\sigma(t)}],\nonumber \\
H & = & p\cdot\dot{x}-L=\frac{1}{2}\frac{m}{\sigma(t)}[\dot{x}^{2}\pm\nu^{2}(x^{2}-t^{2}\dot{x}^{2})\mp2\frac{\nu^{2}x^{2}}{\sigma(t)}]\nonumber \\
 & = & \frac{1}{2}m[(\frac{p}{m}\mp\frac{\nu^{2}tx}{\sigma(t)})^{2}\mp\frac{\nu^{2}(1\pm\nu^{2}t^{2})x^{2}}{\sigma(t)^{2}}]\nonumber \\
 & = & \frac{p^{2}}{2m}\mp\frac{\nu^{2}tx\cdot p}{\sigma(t)}\mp\frac{m\nu^{2}x^{2}}{2\sigma(t)^{2}}.\label{Hamiltonian}\end{eqnarray}
Roughly, the NH Schrödinger equation (\ref{Sch_eq}) can be obtained
from the canonical quantization of the above Hamiltonian, while we
will come back to the subtle problem of operator ordering later. The
corresponding canonical equations of motion are\begin{eqnarray*}
\dot{x}^{i} & = & \{x^{i},H\}=\frac{p_{i}}{m}\mp\frac{\nu^{2}tx^{i}}{\sigma(t)},\\
\dot{p}_{i} & = & \{p_{i},H\}=\pm\frac{\nu^{2}tp_{i}}{\sigma(t)}\pm\frac{m\nu^{2}x^{i}}{\sigma(t)^{2}},\end{eqnarray*}
which clearly leads to\[
\ddot{x}^{i}=\frac{\dot{p}_{i}}{m}\mp\frac{d}{dt}[\frac{\nu^{2}t}{\sigma(t)}]x^{i}\mp\frac{\nu^{2}t}{\sigma(t)}\dot{x}^{i}=0\]
after a bit of calculation. The same EOM $\ddot{x}^{i}=0$ can, of
course, be obtained directly as the Euler-Lagrange equation from the
Lagrangian (\ref{Lagrangian}).

Having the above Lagrangian formalism, it is easy to discuss the NH
quantum mechanics, especially the realization of the NH symmetry in
the quantum case from a path integral point of view. Formally, the
propagator (Feynman kernel) from the space-time point
$(t_{1},x_{1})$ to $(t_{2},x_{2})$ is given by the path integral\[
K(t_{2},x_{2};t_{1},x_{1})=\int
e^{\frac{\mathrm{i}}{\hbar}S[x(t)]}D_{t_{1},x_{1}}^{t_{2},x_{2}}x(t),\]
where $D_{t_{1},x_{1}}^{t_{2},x_{2}}x(t)$ means the appropriate
functional integral measure for functions $x(t)$ satisfying
$x(t_{1})=x_{1}$ and $x(t_{2})=x_{2}$. Take the NH space
translation\begin{eqnarray}
t^{\prime} & = & t,\nonumber\\
x^{\prime i} & = & x^{i}-a^{i}\label{NH_space_trans}\end{eqnarray}
as an example. This transformation acts nontrivially on $S[x(t)]$
as\begin{eqnarray*}
S[x(t)]\to S^{\prime}[x^{\prime}(t^{\prime})] & = & \int_{t_{1}}^{t_{2}}\frac{1}{2}\frac{m}{\sigma(t)}[\dot{x}^{\prime2}\mp\nu^{2}(x^{\prime}-t\dot{x}^{\prime})^{2}\pm2\frac{\nu^{2}x^{\prime2}}{\sigma(t)}]dt\\
 & = & \int_{t_{1}}^{t_{2}}\frac{1}{2}\frac{m}{\sigma(t)}[\dot{x}^{2}\mp\nu^{2}(x-a-t\dot{x})^{2}\pm2\frac{\nu^{2}(x-a)^{2}}{\sigma(t)}]dt\\
 & = & S[x(t)]+\int_{t_{1}}^{t_{2}}\frac{1}{2}\frac{m}{\sigma(t)}[\pm2\nu^{2}a\cdot(x-t\dot{x})\mp\nu^{2}a^{2}\mp4\frac{\nu^{2}a\cdot x}{\sigma(t)}\pm2\frac{\nu^{2}a^{2}}{\sigma(t)}]dt\\
 & = & S[x(t)]+\int_{t_{1}}^{t_{2}}d\left(\mp\frac{m\nu^{2}ta\cdot x}{\sigma(t)}\pm\frac{m\nu^{2}ta^{2}}{2\sigma(t)}\right)\\
 & = & S[x(t)]+\left(\mp\frac{m\nu^{2}ta\cdot x}{\sigma(t)}\pm\frac{m\nu^{2}ta^{2}}{2\sigma(t)}\right)_{t_{1}}^{t_{2}},\end{eqnarray*}
which leads to\begin{eqnarray}
 &  & K(t_{2},x_{2};t_{1},x_{1})\to\nonumber \\
 &  & K^{\prime}(t_{2}^{\prime},x_{2}^{\prime};t_{1}^{\prime},x_{1}^{\prime})=\int e^{\frac{\mathrm{i}}{\hbar}S^{\prime}[x^{\prime}(t^{\prime})]}D_{t_{1},x^{\prime}_{1}}^{t_{2},x^{\prime}_{2}}x^{\prime}(t^{\prime})\label{K_trans}\\
 & = & \exp\frac{\mathrm{i}}{\hbar}\left[\left(\mp\frac{m\nu^{2}t_{2}a\cdot x_{2}}{\sigma(t_{2})}\pm\frac{m\nu^{2}t_{2}a^{2}}{2\sigma(t_{2})}\right)-\left(\mp\frac{m\nu^{2}t_{1}a\cdot x_{1}}{\sigma(t_{1})}\pm\frac{m\nu^{2}t_{1}a^{2}}{2\sigma(t_{1})}\right)\right]K(t_{2},x_{2};t_{1},x_{1}).\nonumber \end{eqnarray}
Combined with the basic property
\begin{equation}\label{Huygens}
\psi(t_{2},x_{2})=\int
K(t_{2},x_{2};t_{1},x_{1})\psi(t_{1},x_{1})d^{d}x_{1}
\end{equation}
of the propagator, the NH transformation (\ref{K_trans}) immediately
gives the transformed wave function\[
\psi^{\prime}(t^{\prime},x^{\prime})=\exp\frac{\mathrm{i}}{\hbar}\left(\mp\frac{m\nu^{2}ta\cdot
x}{\sigma(t)}\pm\frac{m\nu^{2}ta^{2}}{2\sigma(t)}\right)\psi(t,x)\]
under the NH space translation, which is the same as in \cite{NH}.
Other kinds of NH transformations on $\psi(t,x)$ can be similarly
obtained, with the same results as in \cite{NH}, but in a more
systematic and straightforward way. Note that for the NH space
translation (\ref{NH_space_trans}), the path-integral measure in
(\ref{K_trans}), as well as the measure $d^{d}x_{1}$ in
(\ref{Huygens}), is obviously invariant, but for the NH time
translation there is possibly a real factor from the nontrivial
transformation of the path-integral measure, which cannot be
obtained in the above simple discussion. Nevertheless, the full NH
time translation of $\psi(t,x)$, and so the nontrivial
transformation of the path-integral measure, can be determined by
the NH invariance of the Schrödinger equation (\ref{Sch_eq}), as has
been done in \cite{NH}. Considering the corresponding infinitesimal
transformations, it is easy to check that this quantum system
carries a realization of the extended NH algebra (\ref{ext_NH}).

The EOM $\ddot{x}^{i}=0$ is the same as that of a free point
particle in the Galilei space-time, which can be given by the simple
Lagrangian\begin{equation}
L=\frac{1}{2}m\dot{x}^{2},\label{free}\end{equation} so it may be
expected that the complicated Lagrangian (\ref{Lagrangian}) is
equivalent to the above simple one, with the understanding that the
latter is an expression under the Beltrami coordinates in the NH
space-times. In fact, that is the case. We have\begin{equation}
\frac{1}{2}\frac{m}{\sigma(t)}[\dot{x}^{2}\mp\nu^{2}(x-t\dot{x})^{2}\pm2\frac{\nu^{2}x^{2}}{\sigma(t)}]dt=\frac{1}{2}m\dot{x}^{2}dt+d\left(\pm\frac{m\nu^{2}tx^{2}}{2\sigma(t)}\right),\end{equation}
so these two forms of Lagrangian are equivalent up to total
derivatives. According to the above discussion from the path
integral point of view, discarding this boundary term corresponds to
a unitary transformation\begin{equation}
\psi(t,x)=\exp\frac{\mathrm{i}}{\hbar}\left(\pm\frac{m\nu^{2}tx^{2}}{2\sigma(t)}\right)\tilde{\psi}(t,x)\label{psi_trans}\end{equation}
of the wave function. On the one hand, canonical quantization of the
simple Lagrangian (\ref{free}) leads to the standard Schrödinger
equation\begin{equation}
\mathrm{i}\hbar\partial_{t}\psi(t,x)=-\frac{\hbar^{2}\nabla^{2}}{2m}\psi(t,x)\label{standard}\end{equation}
for a free point particle, without any ambiguity. On the other hand,
a direct transformation (\ref{psi_trans}) of the wave function in
(\ref{Sch_eq}) gives\[
\mathrm{i}\hbar\partial_{t}\tilde{\psi}(t,x)=\left(-\frac{\hbar^{2}\nabla^{2}}{2m}\mp\frac{\mathrm{i}\hbar\nu^{2}td}{2\sigma(t)}\right)\tilde{\psi}(t,x),\]
which can be transformed into the standard one (\ref{standard})
through\[ \tilde{\psi}(t,x)=\sigma(t)^{d/4}\psi(t,x).\] Here recall
that $d$ is the space dimension. The above transformation is,
however, not unitary, so it is nontrivial in physics. In fact, if we
take the usual symmetrization prescription for the ordering of
$x^{i}$ and $p_{i}$ to render Hermiticity of the canonical
quantization of the Hamiltonian (\ref{Hamiltonian}) in the usual
sense, we will obtain the Schrödinger equation\[
\mathrm{i}\hbar\partial_{t}\psi(t,x)=[-\frac{\hbar^{2}\nabla^{2}}{2m}\pm\frac{\mathrm{i}\hbar\nu^{2}tx^{i}\partial_{i}}{\sigma(t)}\pm\frac{\mathrm{i}\hbar\nu^{2}td}{2\sigma(t)}\mp\frac{m\nu^{2}x^{2}}{2\sigma^{2}(t)}]\psi(t,x),\]
which becomes the same form as (\ref{standard}) but with $\psi$
replaced by $\tilde{\psi}$:\begin{equation}
\mathrm{i}\hbar\partial_{t}\tilde{\psi}(t,x)=-\frac{\hbar^{2}\nabla^{2}}{2m}\tilde{\psi}(t,x)\label{extra}\end{equation}
upon the unitary transformation (\ref{psi_trans}).

Those two Schrödinger equations (\ref{Sch_eq}) and (\ref{extra})
are not trivially equivalent in physics, and the former has been discussed
in detail in \cite{NH} with its NH invariance and conservation of
probability shown, which justifies it as the correct equation for
the NH quantum mechanics, so we call the quantum mechanics described
by the latter the extraordinary NH quantum mechanics. Nevertheless,
if we allow non-unitary transformations of the wave function, these
two equations are essentially equivalent. In fact, collecting the
results of the above discussion, we see the following transformation\begin{equation}
\psi(t,x)=\sigma(t)^{d/4}\exp\frac{\mathrm{i}}{\hbar}\left(\pm\frac{m\nu^{2}tx^{2}}{2\sigma(t)}\right)\tilde{\psi}(t,x)\label{total_trans}\end{equation}
of wave functions in these two equations. Having this explicit form
of transformation, the NH invariance of (\ref{extra}) is evident,
with the NH transformation properties listed as follows:
\begin{itemize}
\item NH space translation $x^{\prime i}=x^{i}-a^{i}$: $\tilde{\psi}^{\prime}(t^{\prime},x^{\prime})=\tilde{\psi}(t,x)$.
\item NH time translation $t^{\prime}=\frac{t-a^{t}}{\sigma(a^{t},t)}$:
$\tilde{\psi}^{\prime}(t^{\prime},x^{\prime})=\frac{\sigma(a^{t},t)^{d/2}}{\sigma(a^{t})^{d/4}}\exp\frac{\mathrm{i}}{\hbar}\left(\pm\frac{m\nu^{2}a^{t}x^{2}}{2\sigma(a^{t},t)}\right)\tilde{\psi}(t,x)$.%
\footnote{See Appendix \ref{time_trans} for details.%
}
\item NH boost $x^{\prime i}=x^{i}-u^{i}t$: $\tilde{\psi}^{\prime}(t^{\prime},x^{\prime})=\exp\frac{\mathrm{i}}{\hbar}(-mu\cdot x+\frac{1}{2}mu^{2}t)\tilde{\psi}(t,x)$.
\item NH space rotation $x^{\prime i}=O_{\ j}^{i}x^{j}$: $\tilde{\psi}^{\prime}(t^{\prime},x^{\prime})=\tilde{\psi}(t,x)$.
\end{itemize}
The NH space translation, boost and space rotation of the wave
function has exactly the same form as the Galilei ones, which can
also be easily obtained from the path integral point of view for the
simple Lagrangian (\ref{free}).

Note that the density of probability in the (ordinary) NH quantum
mechanics is defined as \cite{NH}\[
\rho=\sigma(t)^{-d/2}\psi^{*}\psi,\]
and then, accordingly, the inner product should be defined as%
\footnote{The generator $\hat{H}=\mathrm{i}\sigma(t)\partial_{t}\mp\mathrm{i}\nu^{2}tx^{i}\partial_{i}=\mathrm{i}\partial_{\tau}$
of the NH time translation is Hermitian under this inner product.%
}\[ \left\langle \psi,\phi\right\rangle
=\int\psi^{*}\phi\sigma(t)^{-d/2}d^{d}x.\] So, given the
transformation (\ref{total_trans}), the corresponding expressions
for $\tilde{\psi}$ are\[
\rho=\tilde{\psi}^{*}\tilde{\psi},\qquad\left\langle
\tilde{\psi},\tilde{\phi}\right\rangle
=\int\tilde{\psi}^{*}\tilde{\phi}d^{d}x,\] which is consistent with
the usual Hermiticity of the canonical quantization of the
Hamiltonian (\ref{Hamiltonian}) mentioned above. These expressions
and the standard form\[
j_{i}=\frac{\mathrm{i}\hbar}{2m}(\tilde{\psi}\partial_{i}\tilde{\psi}^{*}-\tilde{\psi}^{*}\partial_{i}\tilde{\psi})\]
of the flux of probability can also be directly obtained from the
extraordinary NH Schrödinger equation (\ref{extra}). Although $\rho$
and $j_i$ satisfy a form of ``conservation of probability" \cite{NH}
\begin{equation}
\partial_{t}\rho+\partial_i j_i=0
\end{equation}
similar to that in flat space-times, however, the probability
density $\rho$ (as well as the norm of the inner product defined
above) cannot be the physically measurable one, at least in the
standard framework of general relativity, since it is not
NH-invariant. The standard, observable probability density should be
\begin{equation}
\tilde{\rho}=\psi^{*}\psi=\sigma(t)^{d/2}\tilde{\psi}^{*}\tilde{\psi},
\end{equation}
and the NH-invariant inner product is
\begin{equation}
\{,\}=\int\psi^{*}\phi
d^{d}x=\int\tilde{\psi}^{*}\tilde{\phi}\sigma(t)^{d/2}d^{d}x.
\end{equation}

It has long been known that the standard Schrödinger equation (\ref{standard})
or (\ref{extra}) for a free point particle is in fact invariant under
a larger symmetry group than the Galilei group, called the Schrödinger
group \cite{Sch_group}. The additional types of transformations in
the Schrödinger group are the {}``dilatation''\begin{eqnarray}
t^{\prime} & = & D^{2}t,\nonumber \\
x^{\prime i} & = & Dx^{i},\label{dilatation}\\
\psi^{\prime}(t^{\prime},x^{\prime}) & = & D^{-d/2}\psi(t,x),\end{eqnarray}
and the {}``special conformal transformation'' (SCT)\begin{eqnarray}
t^{\prime} & = & \frac{t}{1-kt},\nonumber \\
x^{\prime i} & = & \frac{x^{i}}{1-kt},\label{SCT}\\
\psi^{\prime}(t^{\prime},x^{\prime}) & = & (1-kt)^{d/2}\exp\frac{\mathrm{i}}{\hbar}\left(\frac{mkx^{2}}{2(1-kt)}\right)\psi(t,x),\end{eqnarray}
where the unitary factor $\exp\frac{\mathrm{i}}{\hbar}\left(\frac{mkx^{2}}{2(1-kt)}\right)$
can also be easily read off from the path integral point of view for
the Lagrangian (\ref{free}). In fact, it can be shown that the NH
groups are subgroups of the Schrödinger group. For example, an NH
time translation can be obtained as the combination of an SCT, a (Galilei)
time translation and a dilatation. This situation is just like that
the dS, AdS and Poincaré groups are different subgroups of the conformal
group \cite{conformal}.

The above discussions also enlighten the study of the harmonic oscillator.
It is known that under the static coordinates (\ref{static}) the
NH Schrödinger equation \cite{Gao,NH}\begin{equation}
\mathrm{i}\hbar\partial_{\tau}\psi(\tau,q)=(-\frac{\hbar^{2}\nabla_{q}^{2}}{2m}\mp\frac{1}{2}m\nu^{2}q^{2})\psi(\tau,q)\label{harmonic}\end{equation}
takes the form of the $d$-dimensional anti-harmonic/harmonic oscillator
with an angular frequency $\nu$, which can also be deduced from the
following coordinate transformation:\[
L=\frac{1}{2}\frac{m}{\sigma(t)}[\dot{x}^{2}\mp\nu^{2}(x-t\dot{x})^{2}\pm2\frac{\nu^{2}x^{2}}{\sigma(t)}]=\frac{1}{2}m(\dot{q}^{2}\pm\nu^{2}q^{2})\]
of the Lagrangian. Note that in the above transformation we do not
discard any total derivatives and $\psi(\tau,q)=\psi(t,x)$. From
(\ref{total_trans}), we see that the wave-function transformation
that relates the (anti-)harmonic oscillator (\ref{harmonic}) to the
free point particle (\ref{extra}) is\begin{equation}
\psi(\tau,q)=\begin{cases}
\mathrm{sech}^{d/2}(\nu\tau)\exp\frac{\mathrm{i}}{\hbar}\left(\frac{1}{2}m\nu\tanh(\nu\tau)q^{2}\right)\tilde{\psi}(t,x) & \mbox{(for NH)}\\
\sec^{d/2}(\nu\tau)\exp\frac{\mathrm{i}}{\hbar}\left(-\frac{1}{2}m\nu\tan(\nu\tau)q^{2}\right)\tilde{\psi}(t,x) & \mbox{(for ANH)}\end{cases}\label{static_trans}\end{equation}
in terms of the static coordinates. So each solution of the equation
(\ref{extra}) gives a solution of the equation (\ref{harmonic}),
and vice versa. This fact has been known for some time \cite{Niederer},
but here we deduce it in an elegant geometric way, which makes its
meaning very intuitive. Let us present two examples. The first example
is the plane wave solution of (\ref{extra}), where\[
\tilde{\psi}(t,x)=e^{\frac{\mathrm{i}}{\hbar}(p\cdot x-\frac{p^{2}}{2m}t)}=\begin{cases}
e^{\frac{\mathrm{i}}{\hbar}[p\cdot q\mathrm{sech}(\nu\tau)-\frac{p^{2}}{2m\nu}\tanh(\nu\tau)]} & \mbox{(for NH)}\\
e^{\frac{\mathrm{i}}{\hbar}[p\cdot q\sec(\nu\tau)-\frac{p^{2}}{2m\nu}\tan(\nu\tau)]} & \mbox{(for ANH)}\end{cases}\]
gives\[
\psi(\tau,q)=\begin{cases}
\mathrm{sech}^{d/2}(\nu\tau)\exp\frac{\mathrm{i}}{\hbar}\left(p\cdot q\mathrm{sech}(\nu\tau)-(\frac{p^{2}}{2m\nu}-\frac{1}{2}m\nu q^{2})\tanh(\nu\tau)\right) & \mbox{(for NH)}\\
\sec^{d/2}(\nu\tau)\exp\frac{\mathrm{i}}{\hbar}\left(p\cdot q\sec(\nu\tau)-(\frac{p^{2}}{2m\nu}+\frac{1}{2}m\nu q^{2})\tan(\nu\tau)\right) & \mbox{(for ANH)}\end{cases}.\]
It is straightforward to check that these wave functions do satisfy
(\ref{harmonic}).%
\footnote{It is obvious that even the wave function in the ANH case is not normalizable,
so it is not for a bound state.%
} In fact, it is interesting to notice that in the ANH case the wave
function $\psi(\tau,q)$ is of the same form as the propagator of
the harmonic oscillator, which is a solution of (\ref{harmonic})
by definition. The second example is the ground state solution of
(\ref{harmonic}) for the ANH case, where\[
\psi(\tau,q)=e^{-\frac{m\nu}{2\hbar}q^{2}-\frac{\mathrm{i}d}{2}\nu\tau}=e^{-\frac{m\nu x^{2}}{2\hbar\sigma(t)}-\frac{\mathrm{i}d}{2}\tan^{-1}\nu t}\]
gives\[
\tilde{\psi}(t,x)=\sigma(t)^{-d/4}e^{-\frac{\mathrm{i}d}{2}\tan^{-1}\nu t}\exp\left(\frac{\mathrm{i}}{\hbar}\frac{m\nu^{2}tx^{2}}{2\sigma(t)}-\frac{m\nu x^{2}}{2\hbar\sigma(t)}\right)=(1+\mathrm{i}\nu t)^{-d/2}\exp\left(-\frac{m\nu x^{2}}{2\hbar(1+\mathrm{i}\nu t)}\right).\]
It is again interesting to notice that this wave function is of essentially
the same form as the propagator of the free point particle. Thus,
an interesting duality between the harmonic oscillator and the free
point particle has been shown, which can be investigated in more general
cases.

It is worthwhile to discuss the ANH case, corresponding to the well-behaved
harmonic oscillator, in a little more detail. The most important (Hermitian)
generators are\[
\hat{H}=\mathrm{i}\hbar\partial_{\tau},\qquad\hat{P}_{i}=-\mathrm{i}\hbar\cos(\nu\tau)\frac{\partial}{\partial q^{i}},\qquad\hat{K}_{i}=-\mathrm{i}\nu^{-1}\hbar\sin(\nu\tau)\frac{\partial}{\partial q^{i}}\]
in terms of the static coordinates%
\footnote{The centrally extended version is \[
\hat{P}_{i}=-\mathrm{i}\hbar\cos(\nu\tau)\frac{\partial}{\partial q^{i}}+m\nu x^{i}\sin(\nu\tau),\qquad\hat{K}_{i}=-\mathrm{i}\nu^{-1}\hbar\sin(\nu\tau)\frac{\partial}{\partial q^{i}}-mx^{i}\cos(\nu\tau).\]
}, which satisfy\[
[\hat{H},\hat{P}_{i}]=-\mathrm{i}\hbar\nu^{2}\hat{K}_{i},\qquad[\hat{H},\hat{K}_{i}]=\mathrm{i}\hbar\hat{P}_{i}.\]
In fact, defining the combinations\[
A_{i}=\hat{P}_{i}+\mathrm{i}\nu\hat{K}_{i}=-\mathrm{i}e^{\mathrm{i}\nu\tau}\frac{\partial}{\partial q^{i}},\qquad A_{i}^{\dagger}=\hat{P}_{i}-\mathrm{i}\nu\hat{K}_{i}=-\mathrm{i}e^{-\mathrm{i}\nu\tau}\frac{\partial}{\partial q^{i}},\]
we have\[
[\hat{H},A_{i}]=-\hbar\nu A_{i},\qquad[\hat{H},A_{i}^{\dagger}]=\hbar\nu A_{i}^{\dagger},\]
which clearly shows that $A_{i}$ and $A_{i}^{\dagger}$ act as the
lowering and raising operators, respectively, changing the energies
of stationary states by one level $\hbar\nu$. Thus, the NH group
$N_{-}(1,d)$ acts as something like the dynamical symmetry group
of the harmonic oscillator.

Moreover, since the Schrödinger equation of the free point particle
is invariant under the Schrödinger group, having the above duality
between the (anti-)harmonic oscillator and the free point particle,
we see that the Schrödinger equation (\ref{harmonic}) of the (anti-)harmonic
oscillator is, actually, invariant under the Schrödinger group. There
seems no simple expressions for the finite dilatation and SCT under
the coordinates $(\tau,q^{i})$, so we consider the infinitesimal
version instead. The generator of the dilatation (denoted by $\partial_{D}$)
is of the form\[
2t\partial_{t}+x^{i}\partial_{i}=\begin{cases}
\nu^{-1}\sinh(2\nu\tau)\partial_{\tau}+\cosh(2\nu\tau)q^{i}\frac{\partial}{\partial q^{i}} & \mbox{(for NH)}\\
\nu^{-1}\sin(2\nu\tau)\partial_{\tau}+\cos(2\nu\tau)q^{i}\frac{\partial}{\partial q^{i}} & \mbox{(for ANH)}\end{cases}\]
in terms of $(\tau,q^{i})$. The commutator of $H=\partial_{\tau}$
with $\partial_{D}$ is $[\partial_{\tau},\partial_{D}]=2\partial_{G}$,
where\[
\partial_{G}=\begin{cases}
\cosh(2\nu\tau)\partial_{\tau}+\nu\sinh(2\nu\tau)q^{i}\frac{\partial}{\partial q^{i}} & \mbox{(for NH)}\\
\cos(2\nu\tau)\partial_{\tau}-\nu\sin(2\nu\tau)q^{i}\frac{\partial}{\partial q^{i}} & \mbox{(for ANH)}\end{cases}\]
is actually the combination $2\partial_{t}-\partial_{\tau}=(1\pm\nu^{2}t^{2})\partial_{t}\pm\nu^{2}tx^{i}\partial_{i}$
of the Galilei time translation and the NH time translation%
\footnote{Such kinds of combinations have been discussed in other context \cite{combine}.%
}. It is easy to check that
$[\partial_{\tau},\partial_{G}]=\pm2\nu^{2}\partial_{D}$ and
$[\partial_{G},\partial_{D}]=2\partial_{\tau}$, so we see that these
generators form an $\mathfrak{so}(1,2)$ subalgebra, as required by
the structure of the Schrödinger group. Supplemented with the
generators $\partial_{D}$ and $\partial_{G}$, the extended NH
algebra $\mathfrak{n}_{\pm}^{\mathrm{C}}(1,d)$ becomes the full
Schrödinger algebra. Now we should work out the actions of these
generators on the wave function $\psi(\tau,q)$ using
(\ref{static_trans}), with some important results shown as follows:
\begin{itemize}
\item $\partial_{\tau}$: $\delta\psi=0$.
\item $\partial_{D}$: $\delta\psi=\begin{cases}
\epsilon(-\frac{d}{2}\cosh2\nu\tau+\frac{\mathrm{i}}{\hbar}m\nu q^{2}\sinh2\nu\tau)\psi & \mbox{(for NH)}\\
\epsilon(-\frac{d}{2}\cos2\nu\tau-\frac{\mathrm{i}}{\hbar}m\nu q^{2}\sin2\nu\tau)\psi & \mbox{(for ANH)}\end{cases}$ with $\epsilon$ the infinitesimal parameter.
\item $\partial_{G}$: $\delta\psi=\begin{cases}
\epsilon(-\frac{d}{2}\nu\sinh2\nu\tau+\frac{\mathrm{i}}{\hbar}m\nu^{2}q^{2}\cosh2\nu\tau)\psi & \mbox{(for NH)}\\
\epsilon(\frac{d}{2}\nu\sin2\nu\tau-\frac{\mathrm{i}}{\hbar}m\nu^{2}q^{2}\cos2\nu\tau)\psi & \mbox{(for ANH)}\end{cases}$.
\end{itemize}
It is a little lengthy but straightforward to check the invariance
of (\ref{harmonic}) under these infinitesimal transformations.

\section{Anomalous Newton-Hooke Space-Times\label{sec:Anomalous_NH}}

The standard NH space-times as affine connection spaces have been
discussed in \cite{NH}, with the following nonzero coefficients of
the affine connection:\[
\Gamma_{tt}^{t}=\frac{\pm2\nu^{2}t}{\sigma(t)},\qquad\Gamma_{tj}^{i}=\Gamma_{jt}^{i}=\frac{\pm\nu^{2}t}{\sigma(t)}\delta_{j}^{i}.\]
These coefficients cannot be uniquely determined by the NH-invariant
metrics, since the latter are degenerate. But they can be uniquely
determined by the NH invariance of both the connection itself and
the affine parameter, as shown in the Appendix A of \cite{NH}. In
fact, it is taken for granted in \cite{NH} that the affine parameter
is identified, up to constant linear transformations, to the NH-invariant
proper time $\tau$. However, this identification is too restrictive,
since the affine parameter $\lambda$ is not an observable in general
and need not be NH-invariant. Relaxing this restriction,%
\footnote{For doing this, we actually violate the compatibility between the
affine connection and the (degenerate) metric, as can be checked for
the connection (\ref{connection}), but it seems that no essential
problem will be caused in a theory which is not generally covariant. %
} there can be one arbitrary real parameter $C$, appearing as the
integration constant when solving the first order ODE imposing the
NH invariance, in the nonzero coefficients:\begin{equation}
\Gamma_{tt}^{t}=\frac{\pm2\nu^{2}t+2C\nu}{\sigma(t)},\qquad\Gamma_{tj}^{i}=\Gamma_{jt}^{i}=\frac{1}{2}\Gamma_{tt}^{t}\delta_{j}^{i}=\frac{\pm\nu^{2}t+C\nu}{\sigma(t)}\delta_{j}^{i}.\label{connection}\end{equation}
Now the first integral of the temporal component of the geodesic equation:
\[
\frac{d^{2}t}{d\lambda^{2}}+\Gamma_{\mu\nu}^{t}(t,x)\frac{dx^{\mu}}{d\lambda}\frac{dx^{\nu}}{d\lambda}=0\]
is (omitting the multiplicative integration constant)\[
\frac{dt}{d\lambda}=\sigma(t)\varsigma(t)^{C}\]
with \[
\varsigma(t)=\begin{cases}
\frac{1-\nu t}{1+\nu t} & \mbox{(for NH)}\\
e^{-2\tan^{-1}\nu t} & \mbox{(for ANH)}\end{cases},\]
which means\[
\frac{d\tau}{d\lambda}=\varsigma(t)^{C},\]
and further (omitting the additive integration constant)\[
\lambda=\frac{1}{2C\nu}\varsigma(t)^{-C}-\frac{1}{2C\nu}.\]

We should check whether the NH first law is always respected for arbitrary
$C$. In fact, it is not difficult to prove that the integral of the
full geodesic equation\[
\frac{d^{2}x^{\rho}}{d\lambda^{2}}+\Gamma_{\mu\nu}^{\rho}(t,x)\frac{dx^{\mu}}{d\lambda}\frac{dx^{\nu}}{d\lambda}=0\]
is a straight (world) line for arbitrary $C$. Furthermore, we should
check that $\lambda$ remains affine parameter under NH transformations,
i.e. $\lambda$ transforms linearly, which can be shown as follows:\[
\frac{d\tau}{d\lambda^{\prime}}=\varsigma(t^{\prime})^{C}=\begin{cases}
\left(\frac{1-\nu^{2}a^{t}t-\nu t+\nu a^{t}}{1-\nu^{2}a^{t}t+\nu t-\nu a^{t}}\right)^{C}=\left(\frac{1+\nu a^{t}}{1-\nu a^{t}}\right)^{C}\left(\frac{1-\nu t}{1+\nu t}\right)^{C} & \mbox{(for NH)}\\
e^{-2C(\tan^{-1}\nu t-\tan^{-1}\nu a^{t})}=e^{2C\tan^{-1}\nu a^{t}}e^{-2C\tan^{-1}\nu t} & \mbox{(for ANH)}\end{cases}=\frac{\varsigma(t)^{C}}{\varsigma(a^{t})^{C}}\]
leading to\[
\frac{d\lambda^{\prime}}{d\lambda}=\varsigma(a^{t})^{C}.\]
Thus the space-times with the NH-invariant metric $d\tau^{2}$ and
affine connection (\ref{connection}) should be considered as qualified
NH space-times, which we call the anomalous NH space-times.

It is straightforward to show that the NH-invariant curvature tensor
and Ricci tensor corresponding to the connection (\ref{connection})
have the following nonzero components\begin{equation}
R_{t\mu\nu}^{i}=\pm\frac{(1\mp C^{2})\nu^{2}}{\sigma(t)^{2}}(\delta_{\mu}^{t}\delta_{\nu}^{i}-\delta_{\mu}^{i}\delta_{\nu}^{t}),\qquad R_{tt}=\mp\frac{(1\mp C^{2})\nu^{2}d}{\sigma(t)^{2}}.\label{curvature}\end{equation}
It can be seen that the case of $C=1$ for NH is very special%
\footnote{The discussion for $C=-1$ is similar.%
}, where the curvature tensor vanishes and the space-time becomes totally
flat. In this case, (\ref{connection}) becomes\[
\Gamma_{tt}^{t}=\frac{2\nu}{1-\nu t},\qquad\Gamma_{tj}^{i}=\Gamma_{jt}^{i}=\frac{\nu}{1-\nu t}\delta_{j}^{i}.\]
Since the space-time is flat, there should be a coordinate system
in which the affine connection also vanishes. It can be checked that\begin{eqnarray}
\lambda & = & \frac{1}{2\nu}\frac{1+\nu t}{1-\nu t}-\frac{1}{2\nu}=\frac{t}{1-\nu t},\nonumber \\
y^{i} & = & \frac{x^{i}}{1-\nu t}\label{lambda-y}\end{eqnarray}
is just such a coordinate system (with $-(2\nu)^{-1}<\lambda$). The
coordinates $(\lambda,y^{i})$ are related to $(t,x^{i})$ by a fractional
linear transformation (with common denominator), so a free point particle
still moves along a straight line with uniform velocity in terms of
$(\lambda,y^{i})$. Even if $C$ is arbitrary, $(\lambda,y^{i})$
are interesting and useful coordinates on the NH space-times. In fact,
the realization of the NH transformation under these coordinates can
be worked out as\begin{eqnarray}
\lambda^{\prime} & = & \frac{t^{\prime}}{1-\nu t^{\prime}}=\frac{t-a^{t}}{1-\nu^{2}a^{t}t-\nu t+\nu a^{t}}=\varsigma(a^{t})\lambda-\frac{a^{t}}{1+\nu a^{t}},\nonumber \\
y^{\prime i} & = & \frac{x^{\prime i}}{1-\nu t^{\prime}}=\frac{\sigma(a^{t})^{1/2}O_{\ j}^{i}(x^{j}-a^{j}-u^{j}t)}{1-\nu^{2}a^{t}t-\nu t+\nu a^{t}}=\varsigma(a^{t})^{1/2}O_{\ j}^{i}(y^{j}-a^{j}-w^{j}\lambda)\label{y_trans}\end{eqnarray}
with $w^{j}=u^{j}+\nu a^{j}$, which is a linear transformation instead
of a fractional linear one. Thus we call $(\lambda,y^{i})$ the linear
coordinates. Note that (\ref{lambda-y}) is of the same form as the
SCT (\ref{SCT}), while (\ref{y_trans}) is of the same form as the
dilatation (\ref{dilatation}) if discarding the inhomogeneous terms,
which shows an interesting relationship to the Schrödinger group.
The metric (\ref{metric}) under the linear coordinates is\[
d\tau^{2}=(1+2\nu\lambda)^{-2}d\lambda^{2}.\]

It is also possible to consider the Newton-Cartan-like gravity in
the anomalous NH space-times, similar to the standard case in
\cite{NH}. Rewriting the NH second law\[
m\frac{d^{2}x^{i}}{dt^{2}}=F^{i}\] for the gravitational interaction
as\[
\frac{d^{2}x^{i}}{d\lambda^{2}}-\frac{F^{i}}{m}\frac{dt}{d\lambda}\frac{dt}{d\lambda}+2\Gamma_{tj}^{i}\frac{dt}{d\lambda}\frac{dx^{j}}{d\lambda}=0\]
with $\Gamma_{tj}^{i}$ given in (\ref{connection}), we see that the
Newton-Cartan-like connection should be taken as\[
\Gamma_{tt}^{i}=-\frac{F^{i}}{m}\] in addition to
(\ref{connection}). The above equation is NH invariant due to the
tensor-like property of $\Gamma_{tt}^{i}$ under NH transformations,
since the same discussion as in the Appendix C of \cite{NH} is not
spoilt by the introduction of the parameter $C$. Note that the
nonzero components of the curvature tensor and Ricci tensor
(\ref{curvature}) now become\begin{equation}
R_{ttj}^{i}=-R_{tjt}^{i}=-\partial_{j}\Gamma_{tt}^{i}\pm\frac{(1\mp
C^{2})\nu^{2}}{\sigma(t)^{2}}\delta_{j}^{i},\qquad
R_{tt}=\partial_{i}\Gamma_{tt}^{i}\mp\frac{(1\mp
C^{2})\nu^{2}d}{\sigma(t)^{2}},\label{gravity}\end{equation} before
we consider the gravitational field equation. Taking into account
the natural constraints in \cite{NH}, including the NH invariance,
we can assume the following form of the field equation:\[
R_{tt}=[4\pi G\rho(t,x)\mp(1\mp C^{2})\nu^{2}d]g_{tt},\] where
$g_{tt}=\sigma(t)^{-2}$ is the $tt$ component of the (degenerate)
metric, $G$ the gravitational constant, and the mass density
$\rho(t,x)$ a scalar field under NH transformations. Together with
(\ref{gravity}), the above field equation gives\begin{equation}
\partial_{i}\Gamma_{tt}^{i}=\frac{4\pi G\rho(t,x)}{\sigma(t)^{2}},\end{equation}
so the subsequent discussions are actually the same as in \cite{NH},
which gives the following NH law of gravity:\begin{equation}
\frac{d^{2}x^{i}}{dt^{2}}=-\frac{GM}{\sigma(t)^{1/2}}\frac{x^{i}-X^{i}}{|x-X|^{3}}\label{point-like}\end{equation}
for the case of $d=3$ and the test particle in the gravitational
field of a point-like source with mass $M$ and position $X(t)$.
It is not strange that the parameter $C$ does not play an essential
role in the Newton-Cartan-like gravity, since it is the NH invariance,
not the geometry (affine connection and curvature), that largely determines
the final equation (\ref{point-like}).

In fact, the anomalous structure extends to the Galilei case. Now
the relevant limit is $\nu\to0$ and $C\to\infty$ but $\gamma\equiv\nu C$
held fixed, which leads to the following nonzero coefficients of the
Galilei-invariant affine connection:\[
\Gamma_{tt}^{t}=2\gamma,\qquad\Gamma_{tj}^{i}=\Gamma_{jt}^{i}=\gamma\delta_{j}^{i}\]
and the following nonzero components of the Galilei-invariant curvature
tensor and Ricci tensor\[
R_{t\mu\nu}^{i}=-\gamma^{2}(\delta_{\mu}^{t}\delta_{\nu}^{i}-\delta_{\mu}^{i}\delta_{\nu}^{t}),\qquad R_{tt}=\gamma^{2}d.\]
The integral of the temporal component of the geodesic equation is
($\tau=t$ in this case)\[
\frac{dt}{d\lambda}=e^{-2\gamma t},\qquad\lambda=\frac{1}{2\gamma}e^{2\gamma t},\]
and it is easy to show that the integral of the full geodesic equation
is a straight (world) line for arbitrary $\gamma$. However, similar
to the NH case, since the dynamical equations (for example the Newton's
law of gravity) are largely determined by the Galilei invariance,
it is not clear whether the parameter $\gamma$ can play a role in
physical observations.
\begin{acknowledgments}
The author thanks late Prof. H.-Y. Guo very much for his introduction
of the research field to him. The author would also like to thank
Prof. C.-G. Huang, X.-N. Wu, Z. Xu and B. Zhou for helpful discussions.
This work is partly supported by the National Natural Science Foundation
of China (Grant No. 11075206) and the President Fund of GUCAS.
\end{acknowledgments}
\appendix

\section{Newton-Hooke Time Translation of $\tilde{\psi}(t,x)$\label{time_trans}}

We know \cite{NH} that the wave function $\psi(t,x)$ is invariant
under the NH time translation (\ref{NH_t}), i.e.\begin{equation}
\psi^{\prime}(t^{\prime},x^{\prime})=\psi(t,x).\label{psi_time_trans}\end{equation}
Under the primed reference frame, the wave-function transformation
(\ref{total_trans}) is written as\[
\psi^{\prime}(t^{\prime},x^{\prime})=\sigma(t^{\prime})^{d/4}\exp\frac{\mathrm{i}}{\hbar}\left(\pm\frac{m\nu^{2}t^{\prime}x^{\prime2}}{2\sigma(t^{\prime})}\right)\tilde{\psi}^{\prime}(t^{\prime},x^{\prime}).\]
Now noticing\[
\sigma(t^{\prime})=1\mp\nu^{2}\frac{(t-a^{t})^{2}}{\sigma(a^{t},t)^{2}}=\frac{\sigma(a^{t})\sigma(t)}{\sigma(a^{t},t)^{2}}\]
and that $x^{i}$ transforms as \[
x^{\prime i}=\frac{\sigma(a^{t})^{1/2}}{\sigma(a^{t},t)}x^{i}\]
under the NH time translation, we have\begin{eqnarray*}
\psi^{\prime}(t^{\prime},x^{\prime}) & = & \frac{\sigma(a^{t})^{d/4}\sigma(t)^{d/4}}{\sigma(a^{t},t)^{d/2}}\exp\frac{\mathrm{i}}{\hbar}\left(\pm\frac{m\nu^{2}(t-a^{t})x^{2}}{2\sigma(t)\sigma(a^{t},t)}\right)\tilde{\psi}^{\prime}(t^{\prime},x^{\prime})\\
 & = & \frac{\sigma(a^{t})^{d/4}\sigma(t)^{d/4}}{\sigma(a^{t},t)^{d/2}}\exp\frac{\mathrm{i}}{\hbar}\left(\pm\frac{m\nu^{2}tx^{2}}{2\sigma(t)}\mp\frac{m\nu^{2}a^{t}x^{2}}{\sigma(a^{t},t)}\right)\tilde{\psi}^{\prime}(t^{\prime},x^{\prime}).\end{eqnarray*}
Comparing the above equation with (\ref{total_trans}) and taking
into account (\ref{psi_time_trans}), we then obtain the NH time translation
of $\tilde{\psi}(t,x)$ to be proved.

\end{document}